\documentclass[letterpaper, 10 pt, conference]{ieeeconf}  % Comment this line out if you need a4paper

\IEEEoverridecommandlockouts                              % This command is only needed if 
\usepackage[noEnd=true]{algpseudocodex}  % complementary package for algorithms, determines the 'style' 
\usepackage{algorithm}  % for writing algorithms 
\algrenewcommand\algorithmicrequire{\textbf{Input:}}        % use input rather then require

\usepackage{graphicx}   		% embedding of graphics
\usepackage{math_symbols}   		% math symbols used in thesis 
\usepackage[hidelinks]{hyperref} 	% for embedding urls
\usepackage{tabularx}
\usepackage{ctable} 			% for \specialrule command
\usepackage{siunitx}
\usepackage{microtype} 			% improved text formatting 
\usepackage[font=footnotesize]{caption} % ensures no space between figure and caption
\usepackage{optidef}            	% for writing nice optimization problems 
\PassOptionsToPackage{table}{xcolor} 	% optidef is loading xcolor without table option, so need to pass option  
\usepackage{colortbl}
\usepackage{tikz}

\title{\LARGE \bf
Adaptive Koopman Model Predictive Control of Simple Serial Robots 
}

\author{Adriano del Río$^{1}$ and Christoph Stoeffler$^{1}$% <-this % stops a space
\thanks{$^{1}$German Research Center for Artificial Intelligence, Robert-Hooke-Str. 1,
        28359 Bremen, Germany.    
        {\tt\small \{adriano.del\_rio\_fernandez, christoph.stoeffler\}@dfki.de}}%
}

\begin{document}

\maketitle
\thispagestyle{empty}
\pagestyle{empty}

%%%%%%%%%%%%%%%%%%%%%%%%%%%%%%%%%%%%%%%%%%%%%%%%%%%%%%%%%%%%%%%%%%%%%%%%%%%%%%%%
\begin{abstract}
    Approximating nonlinear systems as linear ones is a common workaround to apply control tools tailored for linear systems. This motivates our present work where we developed a data-driven model predictive controller (MPC) based on the Koopman operator framework,
    allowing the embedding of nonlinear dynamics in a higher dimensional, but linear function space. The controller, termed \textit{adaptive} Koopman model predictive control (KMPC), uses online closed-loop feedback to learn and incrementally update a linear representation of nonlinear system dynamics, without the prior knowledge of a model. Adaptive KMPC differs 
    from most other Koopman-based control frameworks that aim to identify high-validity-range models in advance and then enter closed-loop control without further model adaptations.     
    To validate the controller, trajectory tracking experiments are conducted with 1R and 2R robots under force disturbances and changing model parameters. We compare the controller to classical linearization MPC and Koopman-based MPC without model updates, denoted \textit{static} KMPC. The results show that adaptive KMPC can, opposed to static KMPC, generalize over unforeseen force disturbances and can, opposed to linearization MPC, handle varying dynamic parameters, while using a small set of basis functions to approximate the Koopman operator.               
    
\end{abstract}

%%%%%%%%%%%%%%%%%%%%%%%%%%%%%%%%%%%%%%%%%%%%%%%%%%%%%%%%%%%%%%%%%%%%%%%%%%%%%%%%
\section{INTRODUCTION}
In control systems, distinguishing between linear and nonlinear systems is fundamental. Linear systems are characterized by a proportional input-output relationship, which makes them relatively easier to analyze and control and has led to a wealth of tailored mathematical tools and algorithms~\cite{goodwin_2000}. However, most physical systems, especially those in robotics, are nonlinear, which reflects the complex dynamics encountered in practical applications like robotic manipulators, autonomous vehicles, and industrial automation. 
Commonly, to enable linear model predictive control (MPC)~\cite{rossiter_2003_MPC} of such systems, the nonlinear model is locally linearized, allowing to formulate the underlying problem as a convex quadratic program with the linear dynamics appearing as constraints. This offers a computational advantage over nonlinear MPC~\cite{Gruene_2011}, where a non-convex optimization problem must be solved online. However, the linearization approach generally has limited local validity, which makes long-term predictions less reliable. Moreover, it foremost requires the availability of a model, and its deficiency usually makes parameter identification inevitable. 

The Koopman operator~\cite{Koopman_1931} offers an alternative perspective on dynamical systems by \textit{globally} representing nonlinear dynamics in an infinite-dimensional but \textit{linear} space of observable functions. 
Such a representation would have strong implications for optimal control of nonlinear systems, since it would enable formulating convex quadratic programs without losing of accuracy in long-term predictions due to linearization. However, the infinite-dimensionality of the operator restricts its use for practical control applications. We will not provide any mathematical formulation, since numerous works, such as~\cite{brunton_2022_modern_koopman}, already offer rigorous introductions to the Koopman operator. 
\begin{figure}[t]
    \centering
    \includegraphics[width=0.850\columnwidth]{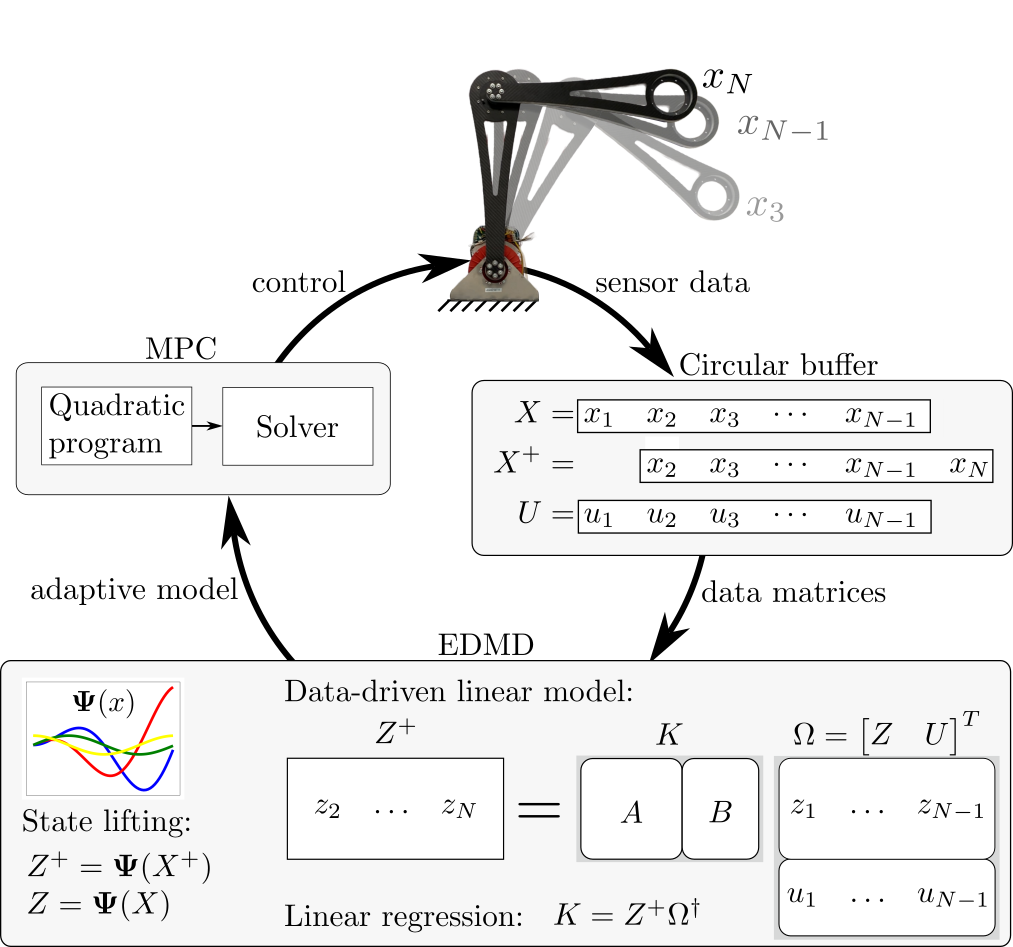}
    \caption{Graphical abstract. In each control cycle, the proposed control architecture relies on sensor feedback, used to update the content of a fixed size databuffer. Extended dynamic mode decomposition is performed recurrently on the data in the buffer, resulting in a data-driven linear model, with validity local to the current operating point. The linear model is used as internal model for a model predictive controller.}\label{fig:paper_concept}
\end{figure}

The emergence of Koopman-based modeling in robotics can be tributed to a set of data-driven algorithms which identify finite-dimensional approximations of the operator, enabling its use for practical control applications~\cite{proctor_2018_DMDc}-\cite{korda_2018_KMPC}. Extended dynamic mode decomposition (EDMD)~\cite{williams_2015_EDMD}, which we address in section~\ref{sec:edmd}, identifies a linear Koopman approximation from data by applying a dictionary of basis functions and solving a linear regression problem. This bypasses modeling from first principles, while granting access to linear control tools as linear MPC. Moreover, real system data can be used~\cite{mamakoukas_2019_local_koopman}-\cite{joglekar_2023_cart}; identified models then inherently capture effects which traditional modeling approaches struggle to include due to complexity or dimensionality.
The quality of the approximation often depends on heuristic selection of basis functions, and their quantity, resulting in a compromise between accuracy and dimensionality of the linear system. Deep learning approaches aim to address these shortcomings by identifying function dictionaries and embeddings directly from data~\cite{Folkestad_2022_koopNet}-\cite{vanderHeijden_2021_DeepKoCo}. This can enhance predictive performance while maintaining computationally tractable number of states.

Particularly challenging in the context of data-driven Koopman modeling are nonlinear systems with multiple fixed points, since their phase space cannot be globally homeomorphic to that of a finite-dimensional linear system~\cite{brunton_2016_invariantsub}. Hence, even deep learning techniques may struggle to find low dimensional, but sufficiently accurate models for systems like the double pendulum~\cite{moore_2024_automatedGA}, which exhibits four fixed points. In such cases, a finite-dimensional Koopman approximation may at most achieve topological equivalence within the entire basin of attraction of a fixed point~\cite{mezic_2013_linearization}. Yet, most other Koopman-based control frameworks attempt to first identify models accurate over the entire phase space, which are then embedded in closed-loop control~\cite{korda_2018_KMPC}-\cite{vanderHeijden_2021_DeepKoCo}. This raises the question: can an alternative approach that re-approximates the Koopman operator based on the current operating point be beneficial for controlling such systems? 

\subsection{Contribution and outline}
In this work, we introduce a control architecture which combines EDMD with linear MPC. In light of aforementioned challenges, we propose an approach where the Koopman operator is recurrently approximated in each control cycle, as schematically depicted in Fig.~\ref{fig:paper_concept}. 
Our control pipeline relies on experimental data that can be gathered without prior model knowledge. We conduct reference tracking experiments with 1R and 2R robot, which exhibit two and four fixed points, respectively, 
and demonstrate the controllers' ability to handle force disturbances and varying dynamic parameters, while using a small set of heuristically chosen basis functions. As baseline for comparison serves linear MPC, which relies on linearization of a nonlinear model, and a Koopman-based controller without model updates. 

The rest of this paper is organized as follows. In section~\ref{sec:prelim}, we introduce the EDMD algorithm and formulate the convex optimization problem solved throughout the control process. In section~\ref{sec:ctrl_synthesis}, we explain the control architecture. Section~\ref{sec:exp} entails a description of experiments we carried out with 1R and 2R robot, followed by a conclusion of our work in section~\ref{sec:conclusion}. 
\section{PRELIMINARY}\label{sec:prelim}
In the following, we use $I_s$ to denote an identity matrix of size $s$, and $0_{a \times b}$ and $1_{a \times b}$ denote matrices with dimensions $a \times b$, filled with zeros and ones,  respectively. When the dimensions are obvious from the context, we will drop the subscripts to improve readability. 

\subsection{Extended dynamic mode decomposition}\label{sec:edmd}
EDMD was introduced in~\cite{williams_2015_EDMD} for autonomous systems and in~\cite{williams_2016_EDMDc} and~\cite{korda_2018_KMPC} extended to controlled systems. In EDMD, $N$~measurements of the state of a system, $x \in \R^n$, are arranged in two time shifted matrices, $\X = \begin{bmatrix} \bxk{1}\; \bxk{2} \cdots \bxk{N-1} \end{bmatrix}$ and $\Xbar = \begin{bmatrix} \bxk{2}\; \bxk{3} \cdots \bxk{N} \end{bmatrix}$, each with dimensions ${n \times {N\!-\!1}}$. The time series data can be from one or several concatenated state trajectories, each sampled at uniform time intervals, $\Delta t$, which have to be small enough to capture the underlying system dynamics. Further, the history of applied controls,~${\bu \in \R^m}$, is organized in matrix $\U =\begin{bmatrix} \buk{1}\; \buk{2} \cdots \buk{N-1} \end{bmatrix}$, with dimensions ${m \times {N\!-\!1}}$. A dictionary of scalar basis functions, 
\vspace{-3mm}
\begin{equation}\label{eq:basis_func_dict}
    \Dict =
    \begin{bmatrix} \dict{1}\;\dict{2} \cdots \dict{p} \end{bmatrix}^{T},
\end{equation}
is applied column wise to the state data matrices to create `lifted' versions of the state vector, which we denote as $\bzk{k} = \Dictof{\bxk{k}} \!\in \!\R^p$. The resulting lifted data matrices are ${\Xlift = \Dictof{\X}}$ and ${\Xbarlift\!= \Dictof{\Xbar}}$, each with dimensions ${p \times N-1}$. 

EDMD assumes that states in the lifted space evolve forward in time in a linear fashion,
and controls are linearly mapped into the lifted state space. This can be expressed as:   
\begin{equation}\label{eq_EDMDc_assumed_relation}
    \Xbarlift = \Aedmd \Xlift + \Bedmd \U 
    = 
    \begin{bmatrix}
        \Aedmd & \Bedmd \\
    \end{bmatrix}
    \begin{bmatrix}
        \Xlift \\
        \U
    \end{bmatrix}
    =
    \Kedmd
    \Omegaedmd,
\end{equation}
with $\Aedmd \in \R^{p \times p}$ and $\Bedmd \in \R^{p \times m}$ being data-driven state transition and control matrix, respectively, and $\Omegaedmd = [\Xlift \U]^T$. $\Kedmd \in \R^{p \times p+m}$ is a finite-dimensional approximation of the Koopman operator, which is determined by solving the linear least squares problem  
\begin{equation}\label{eq:EDMD_problem}
    \begin{aligned}
    & \underset{\Kedmd}{\text{minimize}}
    &&  \norm[\big]{\Xbarlift - \Kedmd \Omegaedmd}_F, 
    \end{aligned}
\end{equation}
where $\norm[]{\cdot}_F$ is the Frobenius norm. The solution is computed as
\begin{equation}\label{eq:K_from_zplus_pinv_omega}
    \Kedmd
    = 
    \begin{bmatrix}
        \Aedmd & \Bedmd \\
    \end{bmatrix}    
    =
    \Xbarlift \Omegaedmd^\dagger,
\end{equation}
where $\Omegaedmd^\dagger$ denotes the Moore-Penrose pseudoinverse of $\Omegaedmd$. The resulting system of equations evolves the lifted state forward in time: 
\begin{equation}\label{eq_EDMDc_lifted_dynamics}
    \bzk{k+1} = \Aedmd \bzk{k} + \Bedmd \buk{k}. 
\end{equation}

To facilitate reconstruction of the original state, the un-lifted state variables, $x$, are included in the dictionary as its first $n$ elements, i.e. $\Dict = \begin{bmatrix} x^T \; \dict{n+1}\cdots \dict{p} \end{bmatrix}^{T}$. The state is then reconstructed as $\bxk{k} 
\!= \!\Cedmd \bzk{k}$, where $\Cedmd \!= \!\begin{bmatrix} I_n\!\!&\!\!\!\!0\end{bmatrix} \in \R^{n \times p}$.

\subsection{Convex model predictive control}\label{sec:convexMPC}
The convex optimization problem sequentially solved in linear MPC relies on a quadratic objective function, $J$, a linear model in discrete-time as equality constraints, and inequality constraints which can represent physical system limits. It writes, using~(\ref{eq_EDMDc_lifted_dynamics}), as  
\begin{mini!}
    { \buk{k},\bzk{k}}{J = \sum_{k=0}^{H-1} \bek{k}^T \Q \bek{k} + \buk{k}^T \Rw \buk{k},\label{eq:opt_obj}}{}{}
    \addConstraint{\!\bzk{k+1}\!}{=\Aedmd \bzk{k} + \Bedmd \buk{k},\;\;\label{eq:opt_con1}}{\!k=0,\ldots,H-1}
    \addConstraint{u_l}{\leq \buk{k} \leq u_u, \quad\label{eq:opt_con2}}{\!k=0,\ldots,H-1}
    \addConstraint{\bzk{0}}{= \Dictof{\bxk{0}}, \quad\label{eq:opt_con3}}{}    
\end{mini!}
where $H$ denotes the prediction horizon and $\bek{k} = \Dictof{\brk{k}} - \bzk{k}$ is the predicted error between the lifted reference $\brk{k}$ and the lifted state. Diagonal matrices $\Q \in \R^{p\times p}$ and $\Rw \in \R^{m\times m}$ define how much emphasis is placed on reference tracking and control effort, respectively. $u_l$ and $u_u$, both $ \in \R^m$, are upper and lower limits on applied controls, and $\Dictof{x_0}$ is the lifted state at the current operating point.  

If the objective includes a term weighting absolute control efforts and the system approaches steady-state, the controller may command inputs, which lead to an offset from the desired reference~\cite{rossiter_2003_MPC}. Intuitively, the error term in (\ref{eq:opt_obj}) becomes zero when the system state coincides with the reference. However, $\buk{k}$ is not necessarily zero at the reference, thus there may be a trade-off solution which is `cheaper'. To avoid this bias, the problem can be expressed in terms of relative controls $\delta u_k  = \buk{k} - \buk{k-1}$, which ensure the optimum corresponds to zero tracking error. We therefore rewrite~(\ref{eq_EDMDc_lifted_dynamics}) as augmented state-space model: 
\begin{equation}\label{eq:lifted_augmented_dynamics}
    \hat{z}_{k+1} = \hat{A} \hat{z}_{k} + \hat{B} \delta u_k, 
\end{equation}  
where augmented state, state transition and control matrix are
\begin{equation*}
    \hat{z}\!:= 
    \begin{bmatrix}
        \bzk{k} \\
        \buk{k-1} \\
    \end{bmatrix}\!, 
    \hat{A}\!:= 
    \begin{bmatrix}
        \A & \B \\
        0 & I_m \\
    \end{bmatrix}\!,   
    \hat{B}\!:= 
    \begin{bmatrix}
        \B\\
        I_m \\
    \end{bmatrix}.        
\end{equation*}
Accordingly, in~(\ref{eq:opt_obj}), we replace the term weighting absolute controls by 
an equivalent term for relative controls. Since the state is now augmented, all state-dependent terms in~(\ref{eq:opt_obj}) and~(\ref{eq:opt_con3}) need to be updated and we hence re-define, by slight abuse of notation,  
\begin{equation*}
    Q=
    \begin{bmatrix}
        Q& 0 \\
        0 &0_{m\times m}\\
    \end{bmatrix}\!
    ,  
    e_k =  
    \begin{bmatrix}
        \Dictof{r_k}\\
        0_{m\times 1}\\
    \end{bmatrix}
    -
    \hat{z}
    ,\,
    \hat{z}_{0} = 
    \begin{bmatrix}
        \Dictof{x_0}\\
        u_{k-1}\\
    \end{bmatrix}.
    % u_l = u_l - u_{k-1}, 
    % u_u = u_u - u_{k-1},     
\end{equation*}
Inequality constraints on the controls are re-expressed in terms of $\delta u$ and we therefore re-write lower and upper bound in (\ref{eq:opt_con2}) as $u_l = u_l - u_{k-1} $ and $ u_u = u_u - u_{k-1}$, respectively. 

Furthermore, it is convenient to translate the optimization problem into a condensed form, as it eliminates state variables from the optimization search space, to make computational effort in the optimization independent from the dictionary size~\cite{korda_2018_KMPC}. This is achieved by expressing future states ${\mathbf{z} = \begin{bmatrix} \hzk{1}^T\cdots \hzk{H}^T\end{bmatrix}^T}$ in terms of the current augmented state 
$\hzk{0}$ and future relative controls ${\mathbf{\delta u} = \begin{bmatrix} \duk{1}^T\cdots\duk{H-1}^T\end{bmatrix}^T}$:
\begin{equation}\label{eq:transform_QP}
    \mathbf{z} = \Apred \hzk{0} + \Bpred \mathbf{\delta u}\; . 
\end{equation}
Here, $\Apred$ is a state transition matrix in block form, and $\Bpred$ is a block lower triangular Toeplitz matrix:
\begin{equation*}
    \Apred\!:= 
    \begin{bmatrix}
        \Ah \\
        \Ah^2 \\
        \vdots \\
        \Ah^{H}
    \end{bmatrix}\!, 
    \Bpred\!:= 
    \begin{bmatrix}
        \Bh & 0 & 0 & \cdots   \\
        \Ah\Bh & \Bh & 0 & \cdots\\
        \vdots & \vdots & \vdots & \ddots\\
        \Ah^{H-1}\Bh & \Ah^{H-2}\Bh & \Ah^{H-3}\Bh & \ddots\\
    \end{bmatrix}.    
\end{equation*}
Both are obtained by iterating~(\ref{eq:lifted_augmented_dynamics}). 
By using (\ref{eq:transform_QP}), the objective can be expressed in general Quadratic Program (QP) form: 
\begin{equation}\label{eq:cost_QP_form}
    J = \mathbf{\delta u}^T (\Bpred^T \mathbf{Q} \Bpred + \mathbf{R}) \mathbf{\delta u} + \mathbf{\delta u}^T
    (2\Bpred^T \mathbf{Q} (\Apred \hzk{0} -\!\mathbf{r})), 
\end{equation}
where ${\mathbf{r}\!=\!\begin{bmatrix}{\Dictof{\brk{1}}^T 0_{1 \times m}}\!\cdots\!\Dictof{\brk{H}}^T 0_{1 \times m} \end{bmatrix}^T}$\!\!\!, with $\Dictof{\brk{k}}$ being the lifted reference states, $\mathbf{Q}\!=\!I_H \otimes Q$, $\mathbf{R}\!=\!I_H \otimes R$, $\otimes$ denotes the Kronecker product. Note that (\ref{eq:opt_con1}) becomes redundant as it is implicitly satisfied. The input constraints become 
% \begin{equation}\label{eq:ineq_new}
    $\boldsymbol{u}_l \leq C_\Delta \delta\mathbf{u} \leq \boldsymbol{u}_u,$ 
% \end{equation}
where $\boldsymbol{u}_l = 1_{H \times 1} \otimes u_l$ and $\boldsymbol{u}_u = 1_{H \times 1} \otimes u_u$. Constraint matrix $C_\Delta = L_1 \otimes I_m$, with $L_1$ being a lower triangular matrix with all entries below and on diagonal set to one. For a detailed derivation, we refer the reader to~\cite{rossiter_2003_MPC}.

\section{CONTROLLER SYNTHESIS}\label{sec:ctrl_synthesis}
\begin{algorithm}[]
    \caption{Adaptive Koopman model predictive control}\label{alg:cap}
    \begin{algorithmic}\label{alg:KMPC}
        % \Function{InterpolatedBufferData}{}
        % \State $P_{u} \gets $  interpolate($t_{k-S:k}, u_{k-S:k}$) \Comment{Polynomial u}
        % \State $P_{x} \gets $ interpolate($t_{k-S:k}, x_{k-S:k}$) \Comment{Polynomial x}
        % \State $t^{'}_{k-n} = t_k - n \Delta t_m$ for $n = 0, \dots, S$ \Comment{S=buffer size}
        % \State $\U \gets $ $P_{u}$($t^{'}_{k-S:k}$) %\Comment{time equidistant controls}
        % \State $\X, \Xbar \gets $ $P_{x}$($t^{'}_{k-S:k-1}$), $P_{x}$($t^{'}_{k-S-1:k}$)         
        % \EndFunction
    
        \Require 
        Reference trajectory: $r_{0:N}$
        \Statex MPC param.: $P = \{Q, R, H, u_l, u_u\}$   \Comment{see (\ref{eq:opt_obj})-(\ref{eq:opt_con3})}
        \Statex Basis functions: $\boldsymbol{\Psi}(x)$ 
        \Statex Preceding experiment data: $\X, \U, T$ \Comment{$T$=time data} 
        \Ensure $k = 0$                     
        \Statex $z_k = \Dictof{x_0}$    
    \State CreateBuffer($\X, \U, T$)                   \Comment{Initially fill buffer}
    \While{$k \leq N$}
    \If{\textit{adaptive scheme} \textbf{or} $k = 0$}
        \State $\X,\Xbar,\U \gets$ \Call{InterpolatedBufferData}{}
        \State $\Xlift,\Xbarlift \gets \Dictof{\X}, \Dictof{\Xbar}$  \Comment{apply~(\ref{eq:basis_func_dict})}    
        \State $\Aedmd,\Bedmd \gets$ LinearRegression($\Xlift,\Xbarlift,\U$) \Comment{see~(\ref{eq:K_from_zplus_pinv_omega})}  
    \EndIf
    \State BuildQP($\Aedmd,\Bedmd, \Dictof{r_{k:k+H}}, P $) \Comment{see section~\ref{sec:convexMPC}}
    \State $u_k \gets$ SolveQP()
    \State $x_k \gets$ ApplySystemControl($u_k$)
    \If{\textit{adaptive scheme}}
        \State UpdateBuffer($x_k, u_k, t_k$)         \Comment{Incremental updates}
    \EndIf
    \State $z_k \gets \Dictof{x_k}$ 
    \State $k \gets k+1$  
    % \State $t \gets t+\Delta t$  
    % \State $t \gets t+\Delta t$
    \EndWhile
    \end{algorithmic}
\end{algorithm}

The combination of Koopman modeling and convex MPC, termed Koopman model predictive control (KMPC), was first proposed in~\cite{korda_2018_KMPC}. There, EDMD is carried out offline, based on 
open loop simulation data. The identified system is then used for online control. We adopt the idea of combining EDMD with convex MPC, but do the following modifications:   
\begin{enumerate}
    \item We restrict data used for the EDMD to come from online experiments, rather than simulations. 
    \item We re-approximate the Koopman operator in each control cycle from recent sensor measurements by integrating EDMD in the online control process.  
\end{enumerate}
Furthermore, we aim to explore how re-approximating the Koopman operator performs in closed-loop control versus determining a Koopman model in advance, as e.g. done in~\cite{korda_2018_KMPC}. We differentiate 
between both approaches by introducing the terms \textit{adaptive} and \textit{static}
KMPC, for recurrent and once-in-advance Koopman operator approximation. 
To facilitate comparison, measured states and applied controls are stored in a circular buffer, operating under \textit{first in, first out} logic. 
To compensate for variations in control frequency, \textit{piecewise cubic hermite polynomials}~\cite{fritsch_2013_hermite} are fitted to the data, and the resulting analytical expression is evaluated at uniform time intervals, which are determined by the mean control frequency. EDMD is then carried out on the time-equidistant data.  
In our adaptive controller, we begin by applying an open-loop sequence of controls,
and use the sensor feedback for building a first linear model. Upon availability, the controller starts tracking a reference; sensor feedback is then used to incrementally update the data buffer. 
Updates are stopped when the system reaches the final goal state within a specified threshold.
In static KMPC, we first use linearization MPC to track a reference trajectory, and then use the measured closed-loop data to obtain a linear model. Once actual reference tracking starts, data buffer updates and EDMD are disabled. We generally denote the process for collecting data at the start as \textit{preceding experiment}. Adaptive and static KMPC are summarized in Algorithm~\ref{alg:cap}.    
As static KMPC uses linearization MPC to collect data at the start, it relies on a model of the system, while the sequence of controls applied in adaptive KMPC before the tracking does not require any model-knowledge. In this regard, adaptive KMPC differs from the Koopman-based online learning approach proposed in~\cite{dittmer_2022} for windfarm control, where a Koopman model is first determined offline based on simulation data, and then updated during closed loop-control. 

\begin{figure}[h]
    \centering
    \includegraphics[width=0.55\columnwidth]{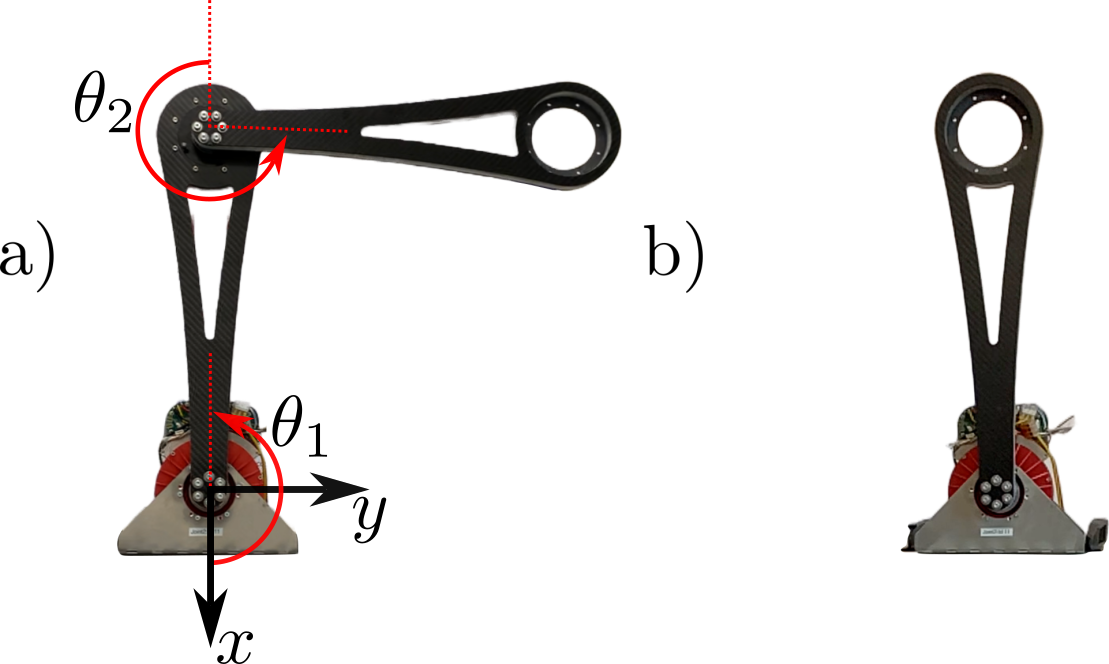}
    \caption{Testbench, set up with 2R robot in a) and 1R robot in b). Both motors are in the red housing, $\theta_2$ in the 2R robot is driven by a belt.}
    \label{fig:testbench}
\end{figure}
\section{EXPERIMENTS}\label{sec:exp}
\subsection{1R and 2R robot testbench}
To validate our approach, we carried out experiments on the testbench shown in Fig.~\ref{fig:testbench}. It consists of fully actuated 2R and 1R robot. Both motor axes coincide with $z$-axis of the base frame and torque transmission to the second joint in a) is achieved via a belt. Both motors exhibit a torque of max.~6\,\si{\newton~\meter}

The inverse dynamic model, which expresses the joint torques, $\bar{u}$, as a function of the generalized joint coordinates $\q$, and their time derivates, in general form is denoted as
\begin{equation}\label{eq:Manipulator_EOM}
    \bar{u} = \Ma\qdd + \Co + \Gr, 
\end{equation}
with $\Ma$ being the generalized inertia matrix, $\Co$ a vector which captures Coriolis and centrifugal forces and $\Gr$ accounting for 
gravity effects. A detailed model of the 2R robot can be found in~\cite{spong_1989_robotics}. We define the state vector of the 1R and 2R robot as $\bx = (\theta_1, \omega_1)$ and $\bx = (\theta_1, \theta_2, \omega_1, \omega_2)$, respectively, with $\theta_i$ being joint angles, computed as illustrated in Fig.~\ref{fig:testbench} a), and $\omega_i$ joint velocities.
\begin{figure*}[t!]
    \centering 
    \includegraphics[width=1.0\textwidth]{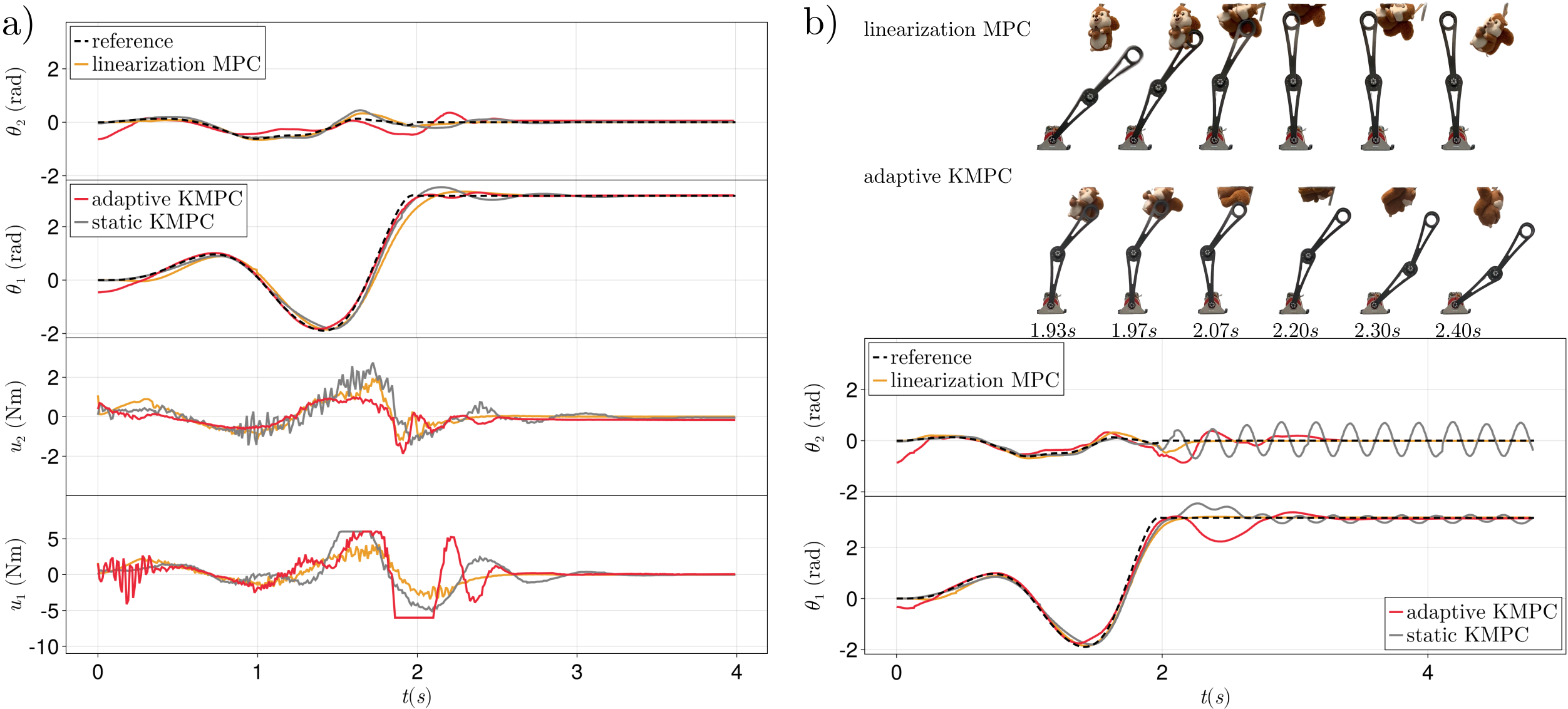}   
    \caption{Experimental results of a) reference tracking and b) force disturbance experiments carried out with the 2R robot. In b) snapshots of the collision between the system and a soft toy hanging from the ceiling are shown at given time intervals.}
    \label{fig:experimental_results}
\end{figure*}
The belt in the 2R system induces a separation of joint and motor space. Hence, joint torques according to~(\ref{eq:Manipulator_EOM}) do not equal motor torques, denoted as $\bu{}$. The motor torques can be computed as $\buk{} = S \bar{u}$, with $S$ being a linear structure matrix. For its derivation, we refer the reader to~\cite{stoeffler_2024_manip}. In our data-driven controllers, we do not account for this mapping, and expect the EDMD to identify the structure matrix implicitly. In our results, we show torques in the motor space.  
\subsection{Reference trajectories}
To generate reference trajectories for tracking experiments, we implemented an iterative linear quadratic regulator (iLQR), according to~\cite{jackson_2019_iLQR}. Trajectories are hence optimized by solving a dynamic programming problem using Bellman's principle of optimality, which requires a dynamic model of the system. However, references can also 
be obtained by model-free methods, e.g.~from simple reference point interpolation.

For all the experiments we used the same trajectory for each system, starting at $x_0 = \begin{bmatrix} 0 \!\!& \!\!0 \!\!&\!\! 0\!\! &\!\!0\end{bmatrix}$ and ending at $x_f = \begin{bmatrix} \pi \!\!& \!\!0 \!\!&\!\! 0\!\! &\!\!0\end{bmatrix}$ for the 2R system, and starting at $x_0 = \begin{bmatrix} 0 \!\!& \!\!0\end{bmatrix}$ and ending at $x_f = \begin{bmatrix} \pi \!\!& \!\!0 \end{bmatrix}$ for the 1R system. This intuitively corresponds to an energy-efficient swing-up to the inverted position. The reference trajectory for the 2R robot is depicted in Fig.~\ref{fig:experimental_results} a) and for the 1R robot in Fig.~\ref{fig:exp_modeluncertainties}. Note, our iLQR implementation does not feature constraints, leading to a reference which exceeds the motor torque threshold. However, we constrain our tracking controllers to remain within physical torque limits using~(\ref{eq:opt_con2}). 
\subsection{Baseline method}
As means of comparison, we use a linearization model predictive controller, which sequentially approximates~(\ref{eq:Manipulator_EOM}) at the current operating point with a first order Taylor series expansion. This allows to use the linear MPC formulation introduced earlier, with slight modifications, which can e.g.~be taken from~\cite{Zhakatayev_2017_slmpc}. 
\subsection{Dictionary functions and controller settings}
Due to the sole presence of trigonometric functions in the equations of motion of both systems, we applied the following lifting functions in the EDMD: 
\begin{equation}
    \Dict= 
    \begin{bmatrix}
    \theta_i & \omega_i & s_i & c_i & \omega_i s_i & \omega_i c_i 
    \end{bmatrix},        
\end{equation}
where $s_i = \sin(\theta_i)$ and $c_i = \cos(\theta_i)$. For the 1R robot, $i = 1$; for the 2R robot $i = 1,2$.  

In static KMPC, we use the same reference trajectory for both, the preceding experiments with the linearization MPC and the subsequent KMPC tracking. In adaptive KMPC, we start by applying a sinusoidal torque sequence as open-loop control signal, inducing oscillations in the system's joint angles $\theta_i$ between $\begin{bmatrix} -0.75\pi,\!\!&\!\!\!0.75\pi \end{bmatrix}$. 
Our choices for $Q$ and $R$ for the controllers can, together with other settings, be found on GitHub\footnote{https://github.com/adrianodelr/adaptive-koopman-mpc}. The testbench revealed to be sensitive to high peak velocities, which seemed to more present in linearization MPC, and weights were used to indirectly enforce avoidance of such. 
Consequently, the controller weights differed during the experiments, however, we ensure comparability by using a performance metric, which evaluates both, used energy and tracking errors. For all experiments the prediction horizon $H$ was set to 30. It must be mentioned, that the underlying communication had notable flaws during the time of experiments and control frequency therefore varied from $90$\,\si{\hertz} to $110$\,\si{\hertz}. 

\subsection{Reference tracking experiments}
\begin{figure}[htbp!]
    \centering
    \includegraphics[width=1.00\columnwidth]{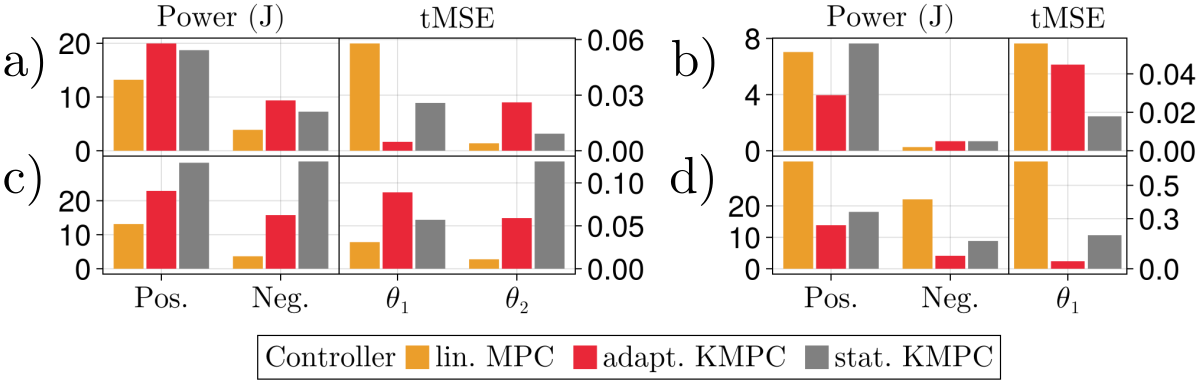}
    \caption{Total positive power used for actuation, total negative power, i.e.~braking energy, both in Joules, and tMSE for the joint angles; shown for all the experiments.}
    \label{fig:energy_tMSE}
\end{figure}

We performed reference tracking experiments with 1R and 2R robot. The results for the 2R robot are depicted in Fig.~\ref{fig:experimental_results} a). The adaptive controller exhibits an offset from the desired reference at the beginning of the tracking process, leading to increased torques. This offset arises from the direct transition from the feedforward torque sequence to the trajectory tracking process. To discard this difference in our performance metrics, we considered trajectories only from 0.75\,\si{\second} onward.   
The power consumption during the experiments, and the time-weighted Mean Squared Error (tMSE) for the joint angles, computed as $1/N \sum_{k=1}^{N} \Delta t_k {(\theta_{i,k}-\tilde{\theta}_{i,k})}^2$, 
are shown in Fig.~\ref{fig:energy_tMSE}~a) and b), for 2R and 1R system, respectively.

Overall, in the 2R experiments, the data-driven controllers exhibited faster and more reactive tracking, at the cost of higher power requirements. This reflects set weights; however, the tMSE for the second joint is higher for the data-driven controllers. 
In the 1R experiments, both data-driven controllers tracked the reference more accurately, with static KMPC consuming slightly more energy and adaptive KMPC consuming less energy than linearization MPC.

\subsection{Force disturbance rejection experiments}
To evaluate how the controllers handle force disturbances, we repeated previously described reference tracking experiments with the 2R robot, but obstructed the reference path with a soft toy hanging from the ceiling, as shown at the top in Fig.~\ref{fig:experimental_results} b). After impact with the soft toy, the linearization controller stabilized the system without notably diverging from the reference trajectory, while adaptive KMPC resulted in more compliant behavior. Static KMPC led to strong oscillations. Power consumption and the tMSE are shown in Fig.~\ref{fig:energy_tMSE} c).
\subsection{Model uncertainty experiments}
\begin{figure}[htbp!]
    \centering
    \includegraphics[width=1.00\columnwidth]{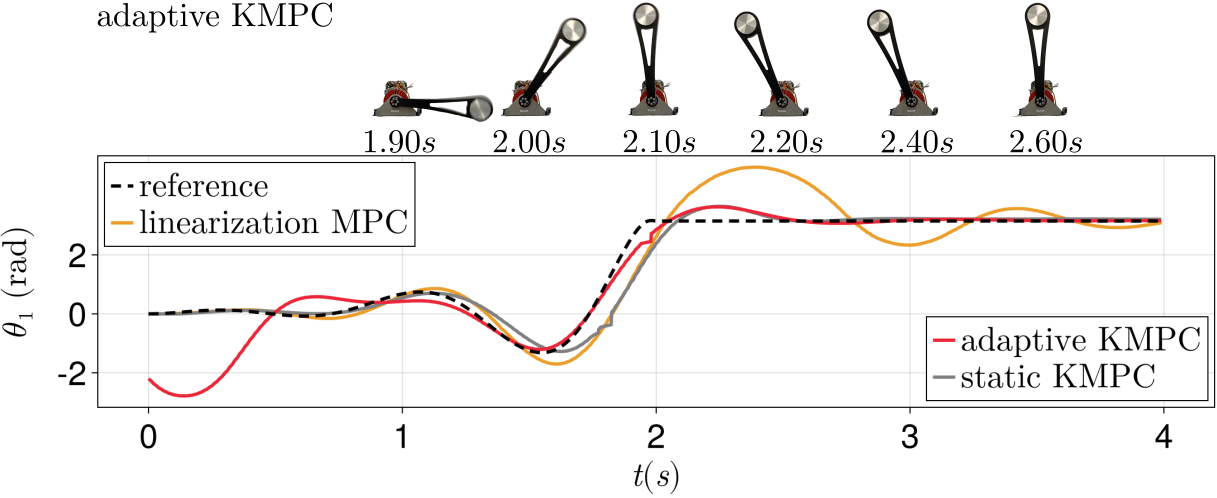}
    \caption{Results of reference tracking experiments with the 1R robot and added end-effector mass. Snapshots show the system before being stabilized by the adaptive controller at the inverted position.}
    \label{fig:exp_modeluncertainties}
\end{figure}
Lastly, we attached a weight of 0.5\,\si{\kilogram} to the end-effector of the 1R robot, without adapting the internal model of the linearization controller. As shown in Fig.~\ref{fig:exp_modeluncertainties}, the linearization MPC strongly overshoots the terminal goal state. Despite rigorous re-tuning, it was not possible to substantially improve tracking. The data-driven controllers led to overshoot, but to a lesser extent. In fact, complete avoidance of the overshoot was not possible, as the breaking torques required to follow the trajectory exceed the torque limit of the motors. Power consumption and tMSE for this experiment are depicted in Fig.~\ref{fig:energy_tMSE} d).

\section{DISCUSSION AND OUTLINE}\label{sec:conclusion}
In this work, we propose a data-driven, convex model predictive controller based on the Koopman operator framework. The control architecture, termed adaptive KMPC, recurrently identifies an internal linear model from online sensor data in real time, without any prior knowledge of the system dynamics. Furthermore, we design a controller, denoted static KMPC, that identifies a linear Koopman model before the control process using online data gathered with a linearization MPC, and consistently solves the convex MPC problem using the same model.

We evaluate the controllers through experiments with 1R and 2R robot, using heuristically chosen dictionaries of basis functions with 6 and 12 dimensions, respectively. We compare their performance against traditional model-based linearization MPC. The experiments show that 
incremental model updates allow adaptive KMPC to generalize over unforeseen force disturbances, 
while static KMPC fails when introduced disturbances are absent from the training data. For static KMPC to generalize over such cases, random input perturbations may be applied when gathering training-data, as done in~\cite{do_2023_KMPC}. Moreover, the data-driven controllers demonstrate superior performance when the analytical model used in linearization MPC is not adapted to changed model parameters, suggesting potential applications of adaptive KMPC in scenarios where dynamic parameters of a system evolve during the control process. 

While this work assessed practical applications of Koopman theory, there remain questions about stability and parameter choices. In \cite{depersis_2020} formalizations for \textit{direct data-driven control}, that shows similarities to our work, were carried out for linear and non-linear systems solely represented by data. In this regard, also more insights about the trajectory buffer size of our controller would need to be gathered to e.g. ensure robustness. Likewise, manual choice of the basis functions for approximating the Koopman operator still relies on some intuition of the underlying system dynamics and has strong implications for the predictive performance of the internal model. Future research directions could hence explore the integration of automated discovery of suitable basis functions,
potentially through the deployment of deep learning techniques. 

\addtolength{\textheight}{-0cm}   % This command serves to balance the column lengths

%%%%%%%%%%%%%%%%%%%%%%%%%%%%%%%%%%%%%%%%%%%%%%%%%%%%%%%%%%%%%%%%%%%%%%%%%%%%%%%%

%%%%%%%%%%%%%%%%%%%%%%%%%%%%%%%%%%%%%%%%%%%%%%%%%%%%%%%%%%%%%%%%%%%%%%%%%%%%%%%%

%%%%%%%%%%%%%%%%%%%%%%%%%%%%%%%%%%%%%%%%%%%%%%%%%%%%%%%%%%%%%%%%%%%%%%%%%%%%%%%%

\section*{ACKNOWLEDGMENT}

The activities described in this paper are part of the project RoLand with support from the Federal Ministry of Food and Agriculture
(BMEL) by decision of the German Bundestag. The Federal Office for Agriculture and Food (BLE) provides coordinating
support for artificial intelligence in agriculture as funding organisation, grant number 28DK103A20 / 28DK103B20 / 28DK103C20.

\end{document}